\documentclass[%
 reprint,
 amsmath,amssymb,
 aps,
prl,
nolongbibliography,
]{revtex4-2}

\usepackage{graphicx}
\usepackage{dcolumn}
\usepackage{bm}
\usepackage{mathrsfs}
\usepackage[dvipsnames]{xcolor}

\usepackage{pdfpages} 
\usepackage{pgffor} 

\makeatletter
\AtBeginDocument{\let\LS@rot\@undefined}
\makeatother

\def\supplementfilename{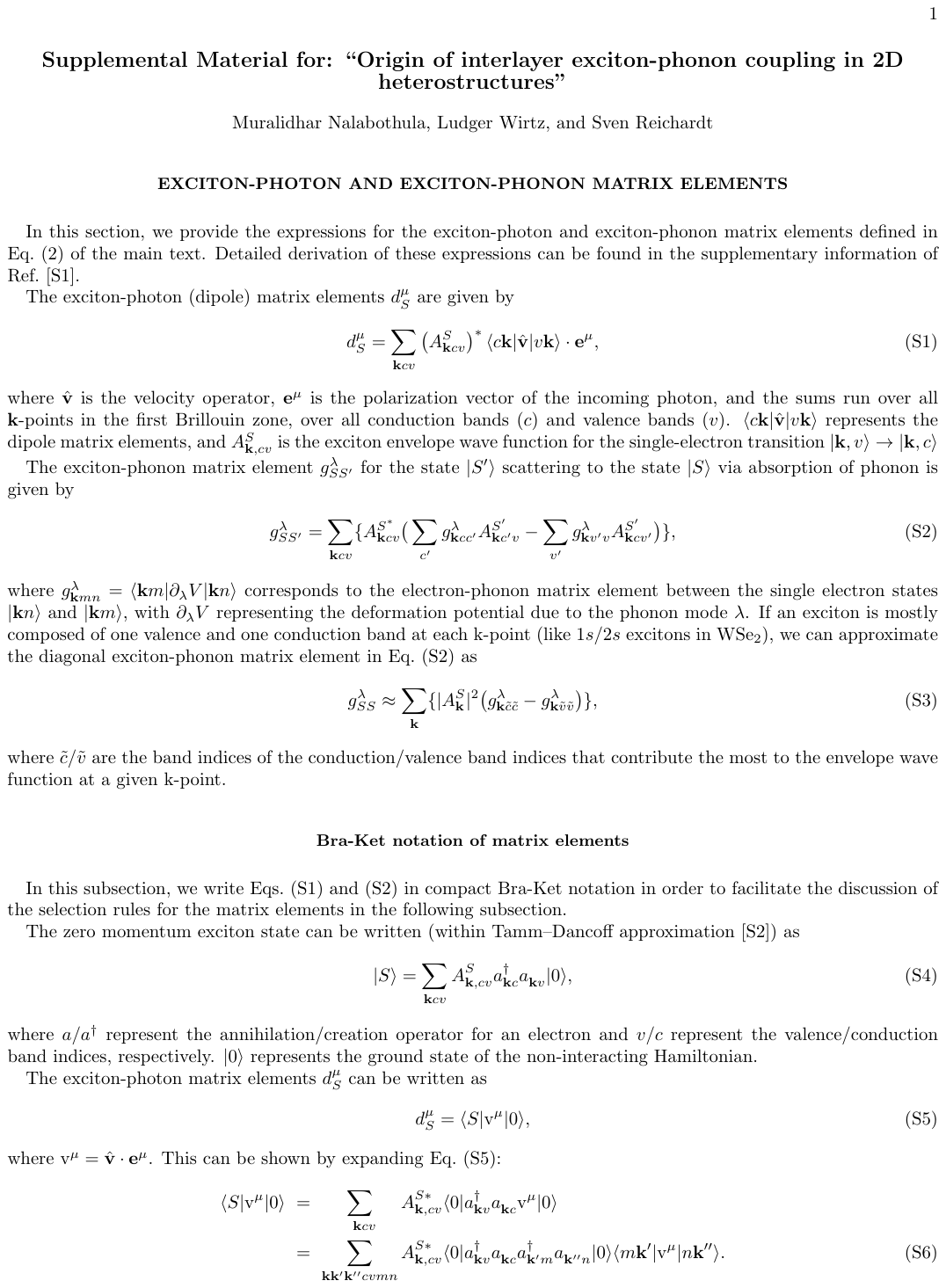}

\pdfximage{\supplementfilename}
\def\numbersupplementpages{\the\pdflastximagepages}

\newif\ifarXiv
\arXivtrue 

\begin{document}

\preprint{APS/123-QED}

\title{Origin of interlayer exciton-phonon coupling in 2D heterostructures}

\author{Muralidhar Nalabothula}
\email{muralidharrsvm7@gmail.com}
\author{Ludger Wirtz}%
\author{Sven Reichardt}%
\affiliation{%
Department of Physics and Materials Science, University of Luxembourg, 162a avenue de la Fa\"iencerie, L-1511 Luxembourg, Luxembourg
}%

\date{\today}

\begin{abstract}

The coupling between excitons and phonons across adjacent layers has been experimentally observed in various heterostructures of layered materials. Yet the precise mechanism underlying this phenomenon remains elusive. Using the WSe$_2$@$h$BN heterostructure as an example, we study the origin of the interlayer exciton-phonon coupling and its signature in resonant Raman scattering through first-principles calculations. Our study emphasizes the central role of crystal symmetries in the interlayer exciton-phonon scattering processes, which are responsible for the anomalous resonant Raman intensities of the in-plane and the out-of-plane $h$BN phonon modes. We find that the deformation potential induced by the $h$BN phonon interacts with the hybridized hole density of WSe$_2$ excitons near the $h$BN interface, leading to interlayer exciton-phonon coupling.
\end{abstract}

\maketitle

In recent years, interfacing two-dimensional (2D) materials with different layers or substrates~\cite{Geim2013} has revealed fascinating properties that are difficult to achieve with individual layers.
For example, interlayer electron-electron interactions, can give rise to diverse phenomena such as moiré excitons~\cite{doi:10.1126/science.aac7820}, superconducting phases~\cite{Cao2018}, and Mott insulating states~\cite{Regan2020}.
Likewise, interlayer electron-phonon interactions can affect the mobilities~\cite{Chen2008} and superconducting temperatures~\cite{wang2012interface} of 2D materials.

Similarly, recent optical scattering measurements on 2D heterostructures have shed light on the coupling of excitons in one layer with phonons of the adjacent layer~\cite{Jin2017,Li2022,ma14020399,PhysRevB.99.205410,PhysRevB.97.235145,Chow2017}.
This remarkable interlayer exciton-phonon coupling was first demonstrated in monolayer WSe$_2$ encapsulated in hexagonal boron nitride ($h$BN) using Raman spectroscopy~\cite{Jin2017}.
Subsequent Raman and photoluminescence measurements have confirmed its existence in various other heterostructures~\cite{Chow2017,ma14020399,Chen2019,Hennighausen2023}.
This interlayer exciton-phonon coupling can play a key role in exciton lifetimes and decoherence times~\cite{Chan2023}, and the study  of phonon polaritons in neighbouring layers~\cite{PhysRevB.104.165404}.

Albeit the experimental evidence for interlayer exciton-phonon coupling, the microscopic mechanism responsible for this phenomenon still remains a mystery. 
In this Letter, we use WSe$_2$@$h$BN as an example to unveil the microscopic mechanism of exciton-phonon coupling across layers. Using \textit{ab initio} methods, we compute resonant Raman intensities, which provide a detailed atomistic view of the interlayer exciton-phonon scattering process. We demonstrate the selection rules in interlayer exciton-phonon scattering, which are responsible for the anomalous resonant Raman intensities: when in resonance with the A exciton of WSe$_2$, the out-of-plane $h$BN phonon mode (which is Raman forbidden in pure $h$BN) exhibits a much higher intensity in the heterostructure than the Raman-allowed in-plane mode~\cite{Jin2017}. Our main findings reveal that the deformation potential of the $h$BN phonon scatters the hybridized part of the WSe$_2$ exciton-hole in the vicinity of the $h$BN layer, giving rise to interlayer exciton-phonon coupling. We show that this coupling is extremely sensitive to the interlayer distance and bond polarity of the phonon layer and is not required at all to observe this effect~\cite{ma14020399,PhysRevB.99.205410,PhysRevB.97.235145,Chow2017}.

We start our discussion by looking at the resonant Raman scattering in a monolayer WSe$_2$@$h$BN heterostructure.
We follow Refs.~\cite{PhysRevB.99.174312, doi:10.1126/sciadv.abb5915} and calculate the differential cross section for Stokes Raman scattering mediated by one phonon:
\begin{equation}
    \frac{\mathrm{d} \sigma}{\mathrm{d} \Omega} \propto \frac{\omega_{\mathrm{L}}-\omega_{\lambda}}{\omega_{\mathrm{L}}}|\mathcal{M}_{\mu\nu}^\lambda(\omega_L,\omega_\lambda)|^2.
\end{equation}
Here, $\mu$ and $\nu$ denote the polarization of the incoming and outgoing light, respectively, while $\omega_{\mathrm{L}}$ and $\omega_{\lambda}$ denote the frequencies of the incoming light and the created phonon of branch $\lambda$, respectively. Within the Tamm-Dancoff approximation~\cite{tda_ref}, the Raman scattering matrix element $\mathcal{M}^{\lambda}_{\mu\nu}$ takes on the simple form
{\small \begin{equation} \label{eq:tda_ram}
    \begin{split}
        \mathcal{M}_{\mu \nu}^\lambda\left(\omega_L, \omega_\lambda\right) = 
        \sum_{S, S^{\prime}} \frac{ \left(d^\nu_{S^\prime}\right)^* (g_{SS^{\prime}}^\lambda)^* d_S^\mu }{\left(\hbar \omega_{\mathrm{L}}-E_S+i \gamma\right)\left(\hbar \omega_{\mathrm{L}}-\hbar \omega_\lambda-E_{S^{\prime}}+i \gamma\right)}  \\
         +\sum_{S, S^{\prime}} \frac{\left(d_S^\mu\right)^* g_{SS^{\prime}}^\lambda d^\nu_{S^{\prime}}}{\left(\hbar \omega_{\mathrm{L}}+E_S-i \gamma\right)\left(\hbar \omega_{\mathrm{L}}-\hbar \omega_\lambda+E_{S^{\prime}}-i \gamma\right)}.
    \end{split}
\end{equation}}
The sums run over all excitonic states $S$ and $S^{\prime}$ with energies $E_{S{(\prime)}}$ and decay constant $\gamma$.
The quantity $d^{\mu}_S$ is the coupling matrix element between an exciton $S$ and a photon of polarization $\mu$, while $(g^{\lambda}_{SS'})^{*}$ represents the exciton-phonon coupling matrix element related to the scattering of an exciton $S$ to an exciton $S'$ \textit{via} emission of one phonon of branch $\lambda$~\cite{doi:10.1126/sciadv.abb5915}.
We obtain all needed quantities for an evaluation of Eq.~\ref{eq:tda_ram} for a hetereostructure of one layer hBN and one layer of WSe$_2$ from first principles using the GW-BSE~\cite{PhysRevB.34.5390,PhysRevB.62.4927} formalism on top of density functional theory (see supplementary information (SM)~\cite{suppli_info} for details of the \textit{ab initio} methods~\cite{PhysRevLett.77.3865,Giannozzi_2017,MARINI20091392,Sangalli_2019,PhysRevB.88.085117,Scherpelz2016,Pizzi_2020,PhysRevB.96.075448,LAZIC2015324,Guandalini2023,PhysRevB.78.085125,PhysRevB.98.245126,10.1145/2427023.2427030,PhysRevLett.62.1169,vesta_citation}).

\begin{figure}[htbp]
    \centering
    \includegraphics[width=\linewidth]{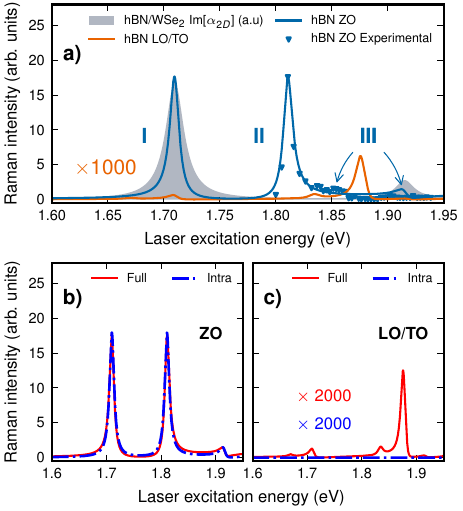}
    \caption{
    Raman intensities of the ZO and LO/TO~modes of $h$BN in a heterostructure of single layers of $h$BN and WSe$_2$ as a function of laser excitation energy.
    a) Computed resonant Raman intensities of the ZO~mode (blue line) and LO/TO~mode (orange line). The gray shaded area denotes the imaginary part of the in-plane polarizability of the heterostructure. Blue triangles represent experimental data of the ZO~mode Raman intensities at 4~K, taken from Ref. \cite{McDonnell_2020}.
    (b) and (c) Computed Raman intensities for (b) the ZO~mode and (c) the LO/TO~mode, considering all scattering channels (red lines) or retaining only intra-exciton scattering channels (blue line).
    }
    \label{fig:raman_fig1}
\end{figure}

In Fig.~\ref{fig:raman_fig1}a, we show the calculated, phonon branch-dependent Raman intensities, as a function of the energy of the incoming photon.
The blue line denotes the Raman intensity for the out-of-plane optical (ZO) phonon of the hBN layer.
We compare the Raman intensity profiles to the imaginary part of the in-plane polarizability (absorption coefficient) of the heterostructure (shaded area) and to experimental Raman intensities from McDonnell~\textit{et al.}~\cite{McDonnell_2020}(blue triangles, see SM~\cite{suppli_info} for further details).
The most striking features of the ZO-mode intensity profile are the two strong resonances at $\omega_{\mathrm{L}} \approx 1.72$~eV and $\omega_{\mathrm{L}} \approx 1.82$~eV, labeled ``I'' and ``II'' in Fig.~\ref{fig:raman_fig1}a.
The first of these two peaks coincides with a resonance in the absorption coefficient which corresponds to the well-known $1s$ state of the ``A''-exciton~\cite{Aslan_2022} of WSe$_2$ and can thus be interpreted as the resonant excitation of the $1s$ A-exciton in WSe$_2$, which then couples to the $h$BN ZO~mode.
The second peak, in contrast, has no such counterpart in the absorption spectrum.
It is the quantum of one ZO~mode away from the first resonance peak and corresponds to the resonant recombination of the $1s$ exciton under phonon emission (when $\omega_{\mathrm{L}} = E_{1s} + \omega_{\mathrm{ZO}}$), compare Eq.~\eqref{eq:tda_ram}.
Finally, Peak~III coincides with the $2s$-A-exciton of the absorption coefficient and is due to the resonant excitation of the $2s$ state.
The results of our calculation match the the experimental data of McDonnell~\textit{et al.}~\cite{McDonnell_2020} well in terms of the local peak intensities.
We note that the $2s$ exciton is blue-shifted in our calculation with respect to experiment, as we consider a heterostructure of one layer WSe$_2$ on top of one layer of $h$BN vs. a system of monolayer WSe$_2$ sandwiched in bulk $h$BN in experiment (where screening is stronger and, thus, the energy difference between $1s$ and $2s$ excitons reduced)~\cite{Aslan_2022}.

Compared to the ZO~mode of $h$BN, its in-plane LO/TO~mode displays a Raman intensity three orders of magnitude lower, yet with a qualitatively similar resonance structure as a function of $\omega_{\mathrm{L}}$.
This remarkable, anomalous behavior of two different allowed phonon modes coupling so differently when the incident light is in resonance with a neighboring layer has been reported in Ref.~\cite{Jin2017}, which provided a speculative interpretation of the underlying Raman scattering process.
Using our atomistic first-principles approach, we can now scrutinize the Raman scattering pathways and understand them in terms of symmetry and involved scattering events.

From Eq.~\eqref{eq:tda_ram}, it is clear that any contributing scattering pathway must feature non-zero optical matrix elements $d^{\mu}_{S}$ and $d^{\nu}_{S'}$ as well as a non-vanishing exciton-phonon coupling matrix element $g^{\lambda}_{S,S'}$.
In order to infer the corresponding selection rules, we note that the underlying $C_3$ point group symmetry of the $h$BN@WSe$_2$ heterostructure allows the classification of zero-momentum phonons and excitons by their total angular momentum component in the out-of-plane direction, with corresponding quantum number $m=+1$, 0, or -1.
A finite optical strength $d^{\mu}_S$ for light polarized parallel to the heterostructure is only possible for excitons $S$ with $m_S=\pm 1$.
Meanwhile, the exciton-phonon matrix elements $g^{\lambda}_{S,S'}$ are non-zero only if total angular momentum (modulo 3) is conserved (see SM~\cite{suppli_info} for derivation):
\begin{equation}
    m_S - m_{S'} - m_\lambda \equiv 0 \text{ (mod 3)}.
    \label{eq:am_conservation}
\end{equation}
For the ZO-phonon ($m_{\mathrm{ZO}}=0$), this implies that active scattering pathways necessarily have $m_S=m_{S'}$.
By contrast, for the LO/TO-phonon ($m_{\mathrm{LO/TO}}=\pm1$), we need to have $m_S = m_{S'} \pm 1$ (mod 3).

The combination of the optical and phonon-specific selection rules allows us then to understand the nature of the resonance features I, II, and III in Fig.~\ref{fig:raman_fig1}a.
As the $1s$ exciton of WSe$_2$ is the only optically active low-energy exciton available, resonance I and II are associated with the scattering process $1s\to1s$, which gives rise to resonances for both the incoming (I) and the outgoing (II) light.
However, while this resonance structure is seen in both the ZO and the LO/TO~mode, the latter is three orders of magnitude less intense, although formerly allowed by symmetry.
To understand this enormous difference between the phonon modes, we first note that the $1s$~exciton of WSe$_2$ is in fact doubly degenerate due to time reversal symmetry~\cite{PhysRevLett.113.026803}.
The two members of the doublet are localized at different, inequivalent corners K and K' of the first Brillouin zone and carry opposite total angular momentum $m_{1s,K^{(\prime)}}=+(-)1$.
For the ZO~mode, this implies that $1s\to1s$ scattering is allowed \emph{within} a valley (\emph{intra}-valley scattering), while for the LO/TO~mode with finite angular momentum, $1s\to1s$ scattering is only allowed between $1s$ states in \emph{opposite} valleys (\emph{inter}-valley scattering).
However, due to the strong localization of the $1s$-exciton wave function in momentum space (see Fig.~S4 of SM~\cite{suppli_info}), there is only a miniscule overlap between wave functions centered in different valleys and, as a result, the inter-valley scattering required for the LO/TO~mode is strongly suppressed and so is consequently the Raman intensity.

In order to substantiate this, we re-calculate the Raman intensities by considering only intra-exciton scattering channels, \textit{i.e.}, by restricting the double sum in Eq.~\eqref{eq:tda_ram} to $S=S'$.
As shown in Fig. \ref{fig:raman_fig1}b and c, we can indeed numerically confirm that intra-exciton scattering is absent for the LO/TO~mode while being the dominant scattering process for the ZO mode.
In terms of the three most prominent resonances in the ZO~mode Raman intensities, we can thus conclude that resonances I and II stem from $1s$-to-$1s$ intra-valley exciton scattering and III from $2s$-to-$2s$ intra-valley scattering.
We note that this finding is in contrast to previous assumptions~\cite{Jin2017} that identified peaks II and III to the two resonances associated with the inter-exciton scattering process $1s\to2s$.
For the LO/TO mode, however, we can confirm that intra-valley scattering plays no role and that the suppressed finite Raman intensity around the $1s$~exciton can only stem from weak inter-exciton scattering.

While these symmetry considerations explain the difference between Raman intensities of the LO/TO- and the ZO-mode, a complimentary analysis is required to understand the precise mechanism for the interlayer exciton-phonon coupling.
In order to elucidate the microscopic mechanism, we consider the $1s\to1s$ scattering pathway for the ZO~mode, which is the dominant Raman scattering channel in the region $\omega_{\mathrm{L}}\lesssim$1.8~eV.
The $1s$ exciton is mostly composed of one conduction and one valence band state~\cite{suppli_info}, therefore we can approximate the $1s \to 1s$ exciton-phonon matrix element as (see SM~\cite{suppli_info} for details)
\begin{equation}
    g_{1s,1s}^{\mathrm{ZO}} \approx \sum_{\mathbf{k}} \left|A^{1s}_{\mathbf{k}cv}\right|^2 \left( g^{ZO}_{\mathbf{k} c} - g^{ZO}_{\mathbf{k} v} \right),
    \label{eq:ex_ph_1s}
\end{equation}
where $A^{1s}_{\mathbf{k}cv}$ is the exciton envelope wave function for the single-electron transition $|\mathbf{k},v\rangle\to|\mathbf{k},c\rangle$.
The diagonal \emph{electron}-phonon matrix elements $g^{\mathrm{ZO}}_{\mathbf{k}n}$ for a single-electron state $|\mathbf{k}n\rangle$ within the framework of DFT are given by~\cite{RevModPhys.89.015003}
\begin{equation}
    g^{\mathrm{ZO}}_{\mathbf{k} n} = \int \mathrm{d}^3\mathbf{r} \, |\psi_{\mathbf{k} n}(\mathbf{r})|^2 \partial_{_\mathrm{ZO}} V_{\text{KS}}(\mathbf{r}).
    \label{eq:elph}
\end{equation}
Here, $\psi_{\mathbf{k}n}(\mathbf{r})$ denotes the one-electron wave function for state $|\mathbf{k}n\rangle$, $V_{\mathrm{KS}}(\mathbf{r})$ corresponds to the total self-consistent Kohn-Sham (KS) potential, and $\partial_{\mathrm{ZO}}$ represents the directional derivative along the ZO phonon mode displacement vector~\cite{RevModPhys.89.015003}.
Combining Eqs.~\eqref{eq:ex_ph_1s} and \eqref{eq:elph}, we obtain
\begin{equation}
    g_{1s,1s}^{\mathrm{ZO}} \approx  \int \mathrm{d}^3\mathbf{r}\, \left[ n^{1s}_c(\mathbf{r}) - n^{1s}_v(\mathbf{r}) \right] 
 \partial_{_\mathrm{ZO}}V_{\mathrm{KS}}(\mathbf{r}),
    \label{eq:ex_ph_1s_welph}
\end{equation}
where $n^{1s}_{c(v)}(\mathbf{r}) = \sum_{\mathbf{k},v(c)} |A^{1s}_{\mathbf{k}cv}|^2 |\psi_{\mathbf{k}c(v)}(\mathbf{r})|^2$ denotes an ``exciton-averaged'' electron ($c$) or hole ($v$) density for the $1s$ exciton.

Previous studies~\cite{ma14020399,PhysRevB.99.205410,PhysRevB.97.235145,Chow2017} have suggested that the polar nature of the ZO phonon is responsible for the origin of the interlayer exciton-phonon coupling.
In order to verify this claim, we separate the total perturbed potential in Eq.~\eqref{eq:ex_ph_1s_welph} into macroscopic and microscopic components~\cite{PhysRevLett.115.176401}:
$\partial_{_\mathrm{ZO}}V_{\mathrm{KS}}(\mathbf{r}) = \partial_{_\mathrm{ZO}}V_{\mathrm{micro}}(\mathbf{r}) + \partial_{_\mathrm{ZO}}V_{\mathrm{macro}}(\mathbf{r})$.
The macroscopic part is the dipole field resulting from the polar nature of $h$BN, \textit{i.e.}, from its Born effective charges, and in the literature is often referred to as the \emph{Fröhlich} field~\cite{PhysRevLett.115.176401}. The microscopic part of the perturbed potential is then calculated by subtracting the macroscopic part from the total perturbed potential.

\begin{figure}[tbp]
    \centering
    \includegraphics[width=\linewidth]{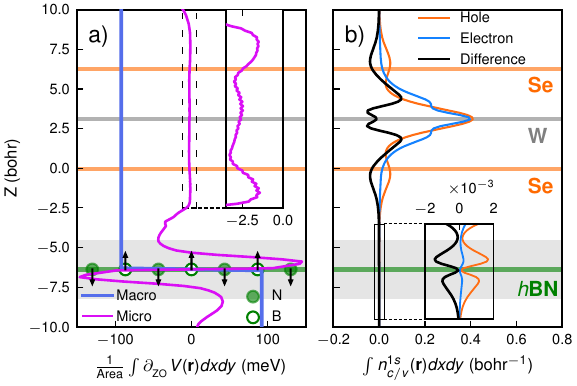}
    \caption{Illustration of the different components of $1s\rightarrow 1s$ exciton-phonon matrix element as given in Eq.~\eqref{eq:ex_ph_1s_welph}, plotted against the out-of-plane spatial coordinate. The horizontal orange, grey, and green lines correspond to the spatial positions of the Se, W, and $h$BN atomic layers, respectively. The grey/white shading represents the near-field/far-field regime of the $h$BN layer.
    (a)  In-plane averaged macroscopic (magenta) and microscopic components (blue) of the total field. The arrows on the B and N atoms represent the ZO phonon mode displacement pattern.
    (b) In-plane integrated electron and hole densities of the $1s$ exciton.
    }
    \label{fig:elecfield_fig2}
\end{figure}

In Figure~\ref{fig:elecfield_fig2}, we illustrate the different components of the $1s\to1s$ exciton-phonon matrix element, as described by Eq.~\eqref{eq:ex_ph_1s_welph}.  We plot the in-plane average of the change in KS potential (divided into microscopic and macroscopic part) and the integrated electron and hole densities as a function of the out-of-plane spatial coordinate ($z$).
We distinguish between the region close to the $h$BN layer, the \emph{near field} (grey shading), and the remaining area, the \emph{far field}.
Figure~\ref{fig:elecfield_fig2}a shows the in-plane average of the microscopic~(magenta line) and macroscopic~(blue line) components of the total field (refer to SM~\cite{suppli_info} for computational details).
The macroscopic component attains a constant value on both sides of the $h$BN layer and extends entirely along the out-of-plane direction, resembling the field generated by a uniformly charged plate.
Conversely, the microscopic component vanishes asymptotically due to the charge neutrality of the system.
Interestingly, the microscopic part exhibits two prominent features: (i) a large and rapidly varying part in the near field, and (ii) a tiny part in the far field that is localized within the WSe$_2$ layer, as depicted in the inset of Fig.~\ref{fig:elecfield_fig2}b. The former arises from the change in lattice potential due to the displacement of nuclei and the induced field of the $h$BN electron density; the latter is due to the induced field generated by the WSe$_2$ electron density~\cite{RevModPhys.89.015003}.

The different components of the total field are felt by the $1s$ exciton charge density of the WSe$_2$ layer.
In Fig.~\ref{fig:elecfield_fig2}b, we depict the in-plane integrated electron and hole densities ($n^{1s}_{c/v}(\mathbf{r})$) of the $1s$ exciton.
Both the electron and hole densities of the $1s$ exciton are mostly made up of tungsten $d$ orbitals~\cite{RevModPhys.90.021001}.
However, while the electron density of the $1s$ exciton is completely localized within the WSe$_2$ layer, the hole density also has a small, but finite component in the $h$BN layer (see inset of Fig.~\ref{fig:elecfield_fig2}c).
This component corresponds to the hybridization of the $p_z$ orbitals of the $h$BN layer with the $d$~orbitals of the WSe$_2$ layer, \textit{i.e.}, to the weak chemical bond between the two layers.
It is noteworthy that the hybridization only occurs for the valence, \textit{i.e.}, bonding orbitals, and not for the conduction band orbitals, owing to the favourable or unfavorable band alignment between the $h$BN and WSe$_2$ valence and conduction bands,  respectively. This hybridization has been experimentally observed and reported in Ref.~\cite{Magorrian_2022}.

\begin{figure}[tbp]
    \centering
    \includegraphics[width=\linewidth]{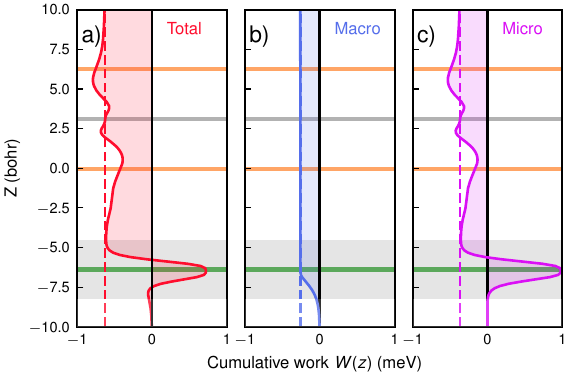}
    \caption{ Cumulative work done by (a) the total field, (b) its macroscopic component, and (c) its microscopic component as defined in Eq.~\eqref{eq:cum_sum}. The vertical dashed lines indicate the asymptotic values.
    }
    \label{fig:cumsum_fig}
\end{figure}

Finally, we consider the interaction of the electron and hole densities of the $1s$ exciton with the total field to gain atomistic insight on the exciton-phonon interaction across layers.
The $1s\rightarrow1s$ exciton-phonon matrix element can \textit{a priori} receive contributions from three different channels:
(i) the constant, macroscopic part interacting with the electron and hole densities over all space (ii) the large near field interacting with the small hybridized hole density in the near-field, and (iii) The small far field in the WSe$_2$ layer interacting with the large electron and hole densities.

To quantify the contribution of each channel to the overall exciton-phonon interaction, we define the cumulative work $W(z)$ along the z axis for a field $\partial V(\mathbf{r})$ as
\begin{equation}
    W(z) = \int_{-\infty}^z \mathrm{d}^3\mathbf{r} \, \big( n^{1s}_c(\mathbf{r}) - n^{1s}_v(\mathbf{r}) \big)(\partial V(\mathbf{r})-\partial V(\mathbf{r}\rightarrow \infty)).
    \label{eq:cum_sum}
\end{equation}
The cumulative work for the total field asymptotically reaches the diagonal exciton-phonon matrix of the $1s$ exciton as given in Eq.~\ref{eq:ex_ph_1s_welph}.

In Fig.~\ref{fig:cumsum_fig}, we depict the cumulative work done by the total field and its constituents, denoting their asymptotic values with dashed, vertical lines.
The cumulative work for all component fields starts from zero and reaches the asymptotic value by the end of the near-field volume.
This implies that the coupling of the $1s$ exciton to the ZO phonon arises almost entirely from the electric fields in the near-field region and their interaction with the hybridized part of the charge density of the $1s$ exciton.
Given that only the hole of the $1s$ exciton hybridizes with the $h$BN orbitals, the $1s \rightarrow 1s$ exciton-phonon scattering can be interpreted as being primarily due to the scattering of the hole by the ZO phonon.

We further examine the contributions of each individual field to the total exciton-phonon coupling by considering the cumulative work of the macroscopic and microscopic components separately.
Although the macroscopic part interacts with the entire electron and hole densities, the cumulative work reaches its asymptotic value already at the interface of the $h$BN layer and the region around the WSe$_2$ layer does not yield a net contribution.
Given that the macroscopic component is nearly constant, the work done by the electron and by the hole compensate each other almost perfectly, as a consequence of the exciton being a charge neutral excitation.
However, the macroscopic part does contribute a finite amount of work to the exciton-phonon coupling through its interaction with the hybridized part of the hole density.
Due to the local horizontal mirror symmetry of the ZO phonon, the macroscopic part flips its sign at the $h$BN layer and consequently does a finite amount of work to the hole (Fig.~\ref{fig:cumsum_fig}a and b).

At the same time, due to lowering of symmetry in the heterostructure, the hybridized hole density around the $h$BN layer is \emph{not} mirror symmetric with respect to the $h$BN plane. In combination with the large microscopic field, this asymmetry leads to a finite amount of work in the near-field region, see Fig.~\ref{fig:cumsum_fig}c.
By contrast, in the far-field regime, the electron and hole distribution are mirror symmetric with respect to the W-layer and in consequence, the net work done by the ZO mode to the exciton cancels out.

It is worthwhile to stress that, unlike the interlayer \emph{electron}-phonon coupling, which is very strong and is predominantly caused by the macroscopic component in the far-field region~\cite{PhysRevMaterials.5.024004}, the interlayer \emph{exciton}-phonon coupling is much weaker and originates in the near-field region, with nearly equal contributions coming from both macroscopic and microscopic components. Remarkably, exciton-phonon scattering takes place even in the absence of the macroscopic component. This implies that bond polarity is not a prerequisite for the occurrence of the interlayer exciton-phonon scattering, contrary to current speculations~\cite{ma14020399,PhysRevB.99.205410,PhysRevB.97.235145,Chow2017}, which hypothesized a macroscopic dipole-dipole coupling mechanism. Furthermore, in Fig.~S5 of the SM~\cite{suppli_info}, we show that the Raman intensities decreases rapidly with increasing inter-layer distance, as observed experimentally in Ref.~\cite{Li2022}, confirming the near-field effect of this phenomenon.

In summary, we theoretically and computationally investigated the coupling of excitons and phonons across layers and its signature in resonant Raman scattering.
Using symmetries, we identified the main exciton-phonon scattering channel in the Raman process of a prototypical WSe$_2$@$h$BN heterostructures.
We find that the \emph{exciton}-phonon coupling is due to the scattering of the hybridized part of the WSe$_2$ exciton hole with the deformation potential of the $h$BN phonon.
This finding sheds light on the nature of the exciton-phonon coupling, which previously was hypothesized to be due to a macroscopic, polar dipole-dipole interaction, but which we find to actually be a near-field effect.
This understanding of the interlayer exciton-phonon coupling in two-dimensional heterostructures can be a valuable guide for tailoring inter-layer interactions in such systems.

\emph{Acknowledgements--}
We acknowledge the use of the HPC facilities of the University of Luxembourg~\cite{VCPKVO_HPCCT22}.
S.R. acknowledges funding from the National Research Fund (FNR) Luxembourg, project C20/MS/14802965/RESRAMAN. L.W. acknowledges funding by the FNR through project C22/MS/17415967/ExcPhon.
We thank Aseem Rajan Kshirsagar and Henry Fried for helpful discussions.

\bibliography{bibilo}

\ifarXiv
    

\section*{Supplementary material}

# Supplemental Material for: "Origin of interlayer exciton-phonon coupling in 2D heterostructures"

Muralidhar Nalabothula, Ludger Wirtz, and Sven Reichardt

## EXCITON-PHOTON AND EXCITON-PHONON MATRIX ELEMENTS

In this section, we provide the expressions for the exciton-photon and exciton-phonon matrix elements defined in Eq. (2) of the main text. Detailed derivation of these expressions can be found in the supplementary information of Ref. [S1].

The exciton-photon (dipole) matrix elements $d_S^\mu$ are given by

$$d_S^\mu = \sum_{\mathbf{k}cv} \left(A_{\mathbf{k}cv}^S\right)^* \langle c\mathbf{k}|\hat{\mathbf{v}}|v\mathbf{k}\rangle \cdot \mathbf{e}^\mu, \tag{S1}$$

where $\hat{\mathbf{v}}$ is the velocity operator, $\mathbf{e}^\mu$ is the polarization vector of the incoming photon, and the sums run over all $\mathbf{k}$-points in the first Brillouin zone, over all conduction bands ($c$) and valence bands ($v$). $\langle c\mathbf{k}|\hat{\mathbf{v}}|v\mathbf{k}\rangle$ represents the dipole matrix elements, and $A_{\mathbf{k},cv}^S$ is the exciton envelope wave function for the single-electron transition $|\mathbf{k},v\rangle \to |\mathbf{k},c\rangle$

The exciton-phonon matrix element $g_{SS'}^\lambda$ for the state $|S'\rangle$ scattering to the state $|S\rangle$ via absorption of phonon is given by

$$g_{SS'}^\lambda = \sum_{\mathbf{k}cv} \{A_{\mathbf{k}cv}^{S*} \big(\sum_{c'} g_{\mathbf{k}cc'}^\lambda A_{\mathbf{k}c'v}^{S'} - \sum_{v'} g_{\mathbf{k}v'v}^\lambda A_{\mathbf{k}cv'}^{S'}\big)\}, \tag{S2}$$

where $g_{\mathbf{k}mn}^\lambda = \langle \mathbf{k}m|\partial_\lambda V|\mathbf{k}n\rangle$ corresponds to the electron-phonon matrix element between the single electron states $|\mathbf{k}n\rangle$ and $|\mathbf{k}m\rangle$, with $\partial_\lambda V$ representing the deformation potential due to the phonon mode $\lambda$. If an exciton is mostly composed of one valence and one conduction band at each k-point (like $1s/2s$ excitons in WSe$_2$), we can approximate the diagonal exciton-phonon matrix element in Eq. (S2) as

$$g_{SS}^\lambda \approx \sum_{\mathbf{k}} \{|A_{\mathbf{k}}^S|^2 \big(g_{\mathbf{k}\tilde{c}\tilde{c}}^\lambda - g_{\mathbf{k}\tilde{v}\tilde{v}}^\lambda\big)\}, \tag{S3}$$

where $\tilde{c}/\tilde{v}$ are the band indices of the conduction/valence band indices that contribute the most to the envelope wave function at a given k-point.

### Bra-Ket notation of matrix elements

In this subsection, we write Eqs. (S1) and (S2) in compact Bra-Ket notation in order to facilitate the discussion of the selection rules for the matrix elements in the following subsection.

The zero momentum exciton state can be written (within Tamm–Dancoff approximation [S2]) as

$$|S\rangle = \sum_{\mathbf{k}cv} A_{\mathbf{k},cv}^S a_{\mathbf{k}c}^\dagger a_{\mathbf{k}v} |0\rangle, \tag{S4}$$

where $a/a^\dagger$ represent the annihilation/creation operator for an electron and $v/c$ represent the valence/conduction band indices, respectively. $|0\rangle$ represents the ground state of the non-interacting Hamiltonian.

The exciton-photon matrix elements $d_S^\mu$ can be written as

$$d_S^\mu = \langle S|\mathrm{v}^\mu|0\rangle, \tag{S5}$$

where $\mathrm{v}^\mu = \hat{\mathbf{v}} \cdot \mathbf{e}^\mu$. This can be shown by expanding Eq. (S5):

$$\begin{aligned}\langle S|\mathrm{v}^\mu|0\rangle &= \sum_{\mathbf{k}cv} A_{\mathbf{k},cv}^{S*} \langle 0|a_{\mathbf{k}v}^\dagger a_{\mathbf{k}c} \mathrm{v}^\mu|0\rangle \\ &= \sum_{\mathbf{k}\mathbf{k}'\mathbf{k}''cvmn} A_{\mathbf{k},cv}^{S*} \langle 0|a_{\mathbf{k}v}^\dagger a_{\mathbf{k}c} a_{\mathbf{k}'m}^\dagger a_{\mathbf{k}''n}|0\rangle \langle m\mathbf{k}'|\mathrm{v}^\mu|n\mathbf{k}''\rangle. \end{aligned} \tag{S6}$$

Employing Wick's theorem for the correlation function, and using $a^\dagger_{\mathbf{k}v}|0\rangle = a_{\mathbf{k}c}|0\rangle = 0$ gives

$$\begin{aligned}\langle S|\mathrm{v}^\mu|0\rangle &= \sum_{\mathbf{k}\mathbf{k}'\mathbf{k}''cvmn} A^{S*}_{\mathbf{k},cv}\delta_{\mathbf{k}''\mathbf{k},vn}\delta_{\mathbf{k}'\mathbf{k},cm}\langle m\mathbf{k}'|\mathrm{v}^\mu|n\mathbf{k}''\rangle \\ &= \sum_{\mathbf{k}cv} A^{S*}_{\mathbf{k},cv}\langle c\mathbf{k}|\mathrm{v}^\mu|v\mathbf{k}\rangle \\ &= d^\mu_S.\end{aligned} \quad (S7)$$

Now, we consider the matrix element $\langle S|\partial_\lambda V|S'\rangle$ between two excitonic states $S'$ and $S$ for phonon absorption. Expanding $|S\rangle$ and $|S'\rangle$ gives

$$\begin{aligned}\langle S|\partial_\lambda V|S'\rangle &= \sum_{\mathbf{k}\mathbf{k}'cc'vv'} A^{S*}_{\mathbf{k}',c'v'}A^{S'}_{\mathbf{k},cv}\langle 0|a^\dagger_{\mathbf{k}'v'}a_{\mathbf{k}'c'}\partial_\lambda V a^\dagger_{\mathbf{k}c}a_{\mathbf{k}v}|0\rangle \\ &= \sum_{\mathbf{k}\mathbf{k}'\tilde{\mathbf{k}}cc'vv'mn} A^{S*}_{\mathbf{k}',c'v'}A^{S'}_{\mathbf{k},cv}\langle 0|a^\dagger_{\mathbf{k}'v'}a_{\mathbf{k}'c'}a^\dagger_{\tilde{\mathbf{k}}m}a_{\tilde{\mathbf{k}}n}a^\dagger_{\mathbf{k}c}a_{\mathbf{k}v}|0\rangle\langle m\tilde{\mathbf{k}}|\partial_\lambda V|n\tilde{\mathbf{k}}\rangle.\end{aligned} \quad (S8)$$

Employing Wick's theorem for the correlation function gives

$$\langle 0|a^\dagger_{\mathbf{k}'v'}a_{\mathbf{k}'c'}a^\dagger_{\tilde{\mathbf{k}}m}a_{\tilde{\mathbf{k}}n}a^\dagger_{\mathbf{k}c}a_{\mathbf{k}v}|0\rangle = -\delta_{\mathbf{k}'\tilde{\mathbf{k}},v'n}\delta_{\mathbf{k}'\mathbf{k},c'c}\delta_{\tilde{\mathbf{k}}\mathbf{k}mv} + \delta_{\mathbf{k}'\mathbf{k},v'v}\delta_{\mathbf{k}'\tilde{\mathbf{k}}c'm}\delta_{\tilde{\mathbf{k}}\mathbf{k}nc} + \delta_{\mathbf{k}'\mathbf{k},v'v}\delta_{\mathbf{k}'\mathbf{k},c'c}\delta_{\tilde{\mathbf{k}}\mathbf{k}mn}. \quad (S9)$$

It is noteworthy to highlight that the last term on the RHS of Eq. (S9) corresponds to a disconnected diagram, and is generally discarded. Moreover, the 3rd term is finite only when $m, n$ belong to valence bands. Substituting Eq. (S9) into Eq. (S8) gives

$$\langle S|\partial_\lambda V|S'\rangle = -\sum_{\mathbf{k}cvv'} A^{S*}_{\mathbf{k},cv'}A^{S'}_{\mathbf{k},cv}g^\lambda_{\mathbf{k}vv'} + \sum_{\mathbf{k}cc'v} A^{S*}_{\mathbf{k},c'v}A^{S'}_{\mathbf{k},cv}g^\lambda_{\mathbf{k}c'c} + \sum_{\mathbf{k}\tilde{\mathbf{k}}cv\tilde{v}} A^{S*}_{\mathbf{k},cv}A^{S'}_{\mathbf{k},cv}g^\lambda_{\tilde{\mathbf{k}}\tilde{v}\tilde{v}}, \quad (S10)$$

where $\tilde{v}$ is a valence band index. Now, on the RHS, we interchange the indices $v$ and $v'$ in the first term and $c$ and $c'$ in the second term. Employing orthogonality of $A_{\mathbf{k}cv}$ in the third term gives

$$\begin{aligned}\langle S|\partial_\lambda V|S'\rangle &= -\sum_{\mathbf{k}cvv'} A^{S*}_{\mathbf{k},cv}A^{S'}_{\mathbf{k},cv'}g^\lambda_{\mathbf{k}v'v} + \sum_{\mathbf{k}cc'v} A^{S*}_{\mathbf{k},cv}A^{S'}_{\mathbf{k},c'v}g^\lambda_{\mathbf{k}cc'} + \delta_{SS'}\sum_{\tilde{\mathbf{k}}\tilde{v}} g^\lambda_{\tilde{\mathbf{k}}\tilde{v}\tilde{v}} \\ &= \sum_{\mathbf{k}cv} A^{S*}_{\mathbf{k},cv}\Big\{\sum_{c'} A^{S'}_{\mathbf{k},c'v}g^\lambda_{\mathbf{k}cc'} - \sum_{v'} A^{S'}_{\mathbf{k},cv'}g^\lambda_{\mathbf{k}v'v}\Big\} + \delta_{SS'}\sum_{\tilde{\mathbf{k}}\tilde{v}} g^\lambda_{\tilde{\mathbf{k}}\tilde{v}\tilde{v}} \\ &= \langle S|\partial_\lambda V|S'\rangle - \delta_{SS'}\sum_{\mathbf{k}v}\langle \mathbf{k}v|\partial_\lambda V|\mathbf{k}v\rangle.\end{aligned} \quad (S11)$$

### DERIVATION OF SELECTION RULES FOR EXCITON-PHONON MATRIX ELEMENTS

Monolayer WSe$_2$ and $h$BN monolayer/bulk individually posses D$_{3h}$ and D$_{3h}$/D$_{6h}$ point group symmetry, respectively. The combined heterostructure still has a reduced C$_3$ point group symmetry, essential for observing valley effects and preserving isotropic properties in these heterostructures [S3].

The zero-momentum exciton and phonon Hamiltonians posses full crystal symmetry. Therefore, the excitonic states and phonon modes of the combined heterostructure transform under representations of the C$_3$ point group. By performing a basis transformation, one can always choose the eigenstates to be irreducible representations. The C$_3$ point group, which is Abelian, posseses three one-dimentional irreducible representations in complex space, denoted as $A$ and $E_\pm$. Therefore, the exciton wave function $|S\rangle$ and the operator $\partial_\lambda V$ (which transforms like a phonon mode) pick up a trivial phase (character) upon the action of point group symmetries:

$$\begin{aligned}\hat{C}_3|S\rangle &= e^{-\frac{2}{3}im_S\pi}|S\rangle \\ \hat{C}_3\,\partial_\lambda V\,\hat{C}_3^\dagger &= e^{-\frac{2}{3}im_\lambda\pi}\partial_\lambda V\end{aligned}$$

where $\hat{C}_3$ is the unitary operator corresponding to 120° rotation about the principal axis which is along the out of plane direction. The allowed values of $m_S$ and $m_\lambda$ are 0 and $\pm 1$ for $A$ and $E_\pm$ representations respectively. Furthermore, we can alternatively define the $\hat{C}_3$ operator as

$$\hat{C}_3 = e^{-\frac{2}{3}i\pi J} \quad (S12)$$

...markdownwhere $J$ is the component of the total angular momentum operator along the principal axis. This implies that $m_S$ and $m_\lambda$ correspond to the total angular momentum of exciton and phonon, respectively, along the principal axis.

Now, we consider the matrix element $\langle S|\partial_\lambda V|S'\rangle$ between the excitonic states $|S\rangle$ and $|S'\rangle$ for the phonon mode $\lambda$:

$$\begin{aligned}\langle S|\partial_\lambda V|S'\rangle &= \langle S|\hat{C}_3^\dagger \hat{C}_3\, \partial_\lambda V\, \hat{C}_3^\dagger \hat{C}_3|S'\rangle \\ &= e^{-\frac{2}{3}i\pi(m_{S'}+m_\lambda-m_S)}\langle S|\partial_\lambda V|S'\rangle.\end{aligned} \qquad (S13)$$

Similarly,

$$\delta_{SS'}\sum_{\mathbf{k}v}\langle \mathbf{k}v|\partial_\lambda V|\mathbf{k}v\rangle = \delta_{SS'}\sum_{\mathbf{k}v}\langle \mathbf{k}v|\hat{C}_3^\dagger \hat{C}_3\, \partial_\lambda V\, \hat{C}_3^\dagger \hat{C}_3|\mathbf{k}v\rangle. \qquad (S14)$$

Application of the $\hat{C}_3$ operator on electronic states gives

$$\hat{C}_3|\mathbf{k}v\rangle = D_{v'v}|\hat{C}_3\mathbf{k},v'\rangle, \qquad (S15)$$

where $\hat{C}_3\mathbf{k}$ is the new k-point obtained by applying 120° rotation to $\mathbf{k}$ and $D_{v'v}$ is a unitary matrix. It follows that

$$\begin{aligned}\delta_{SS'}\sum_{\mathbf{k}v}\langle \mathbf{k}v|\partial_\lambda V|\mathbf{k}v\rangle &= \delta_{SS'}\sum_{\mathbf{k}v}\langle \mathbf{k}v|\hat{C}_3^\dagger \hat{C}_3\, \partial_\lambda V\, \hat{C}_3^\dagger \hat{C}_3|\mathbf{k}v\rangle \\ &= \delta_{SS'}e^{-\frac{2}{3}i\pi m_\lambda}\sum_{\mathbf{k}vv'v''}\langle \hat{C}_3\mathbf{k}v'|\partial_\lambda V|\hat{C}_3\mathbf{k}v''\rangle D^*_{v'v}D_{v''v}.\end{aligned} \qquad (S16)$$

Since $D$ is unitary, $\sum_v D^*_{v'v}D_{v''v}=\delta_{v'v''}$. Changing the indices ($v'\to v$ and $\hat{C}_3\mathbf{k}\to \mathbf{k}$) leads to

$$\begin{aligned}\delta_{SS'}\sum_{\mathbf{k}v}\langle \mathbf{k}v|\partial_\lambda V|\mathbf{k}v\rangle &= e^{-\frac{2}{3}i\pi m_\lambda}\delta_{SS'}\sum_{\mathbf{k}v}\langle \mathbf{k}v|\partial_\lambda V|\mathbf{k}v\rangle \\ &= e^{-\frac{2}{3}i\pi(m_{S'}+m_\lambda-m_S)}\delta_{SS'}\sum_{\mathbf{k}v}\langle \mathbf{k}v|\partial_\lambda V|\mathbf{k}v\rangle.\end{aligned} \qquad (S17)$$

Combining Eq. (S13) and (S17) gives

$$g^\lambda_{SS'} = e^{-\frac{2}{3}i\pi(-m_S+m_\lambda+m_{S'})}g^\lambda_{SS'}. \qquad (S18)$$

The exciton-phonon matrix element $g^\lambda_{SS'}$ is finite only if $m_S - m_{S'} - m_\lambda = 3n$ (where n is an integer) which proves Eq. (3) of the main text. This selection rule represents the discretized version of the conservation of total angular momentum along the principal axis as discussed in the main text.

3

## COMPUTATIONAL DETAILS

All the ground state properties presented in the paper are obtained from density functional theory calculations (DFT) within the generalized gradient approximation [S4] as implemented in the `QUANTUM ESPRESSO` Code [S5]. We use the Perdew-Burke-Ernzerhof (PBE) functional [S4] and fully relativistic norm-conserving pseudo-potentials (SG15 database) to perform all the DFT calculations [S6, S7]. A plane-wave energy cutoff of 120 Ry is used to expand the wave functions. The heterostructure used for all the *ab-initio* calculations is shown in Fig. S1. This heterostructure is generated using the `CELLMATCH` software [S8], with the same parameters as mentioned in Ref.[S9]. We then relax the structure with a convergence threshold of $10^{-5}$ Ry and $10^{-5}$ Ry/Bohr for total energy and forces, respectively. After the relaxation, the obtained interlayer distance between the Tungsten and hBN layers is 5 Å which is in good agreement with Ref. [S9]. In order to avoid spurious effects of out-of-plane periodicity, we set the vacuum separation to 20 Å and employ a Coulomb cutoff in all our *ab-initio* calculations [S10]. We then obtain phonons and electron-phonon coupling matrix elements within density functional perturbation theory as implemented in the `PH.x` code of `QUANTUM ESPRESSO`. A $\Gamma$ centred $9 \times 9 \times 1$ uniform k point grid is used to converge the ground state density and we employ the same k-grid for the phonon calculations.

In order to obtain the total change in the Kohn-Sham potential for plotting purposes, we construct the change in the local part of the ionic potential from the pseudopotentials and then add this to the perturbed Hartree and exchange-correlation potentials obtained from the DFPT calculations. The macroscopic part of the total change in the Kohn-Sham potential is computed using an analytical expression outlined in Ref. [S11].

All the many-body perturbation theory calculations are performed on top of DFT calculations with the `YAMBO` code [S12, S13]. In order to correct the Kohn-Sham band structure, we perform a $G_0W_0$ calculation on a uniform $\Gamma$ centred $12 \times 12 \times 1$ grid with a total of 2600 Kohn-Sham states. A plane wave cutoff of 2 Ry is used for the dielectric tensor to obtain the converged bandgap. The frequency dependence of the dielectric tensor is calculated with the plasmon-pole approximation [S14]. We also use a `G-terminator` [S15] and `RIM-W` technique [S16] to accelerate the convergence of the bandgap with respect to bands and k points respectively. We interpolate the quasi-particle corrections to a finer k-point grid with the `Wannier90` code [S17].

We solve the Bethe–Salpeter equation (BSE) within the Tamm-Dancoff approximation [S2] starting from our previous mean field calculations with the `YAMBO` code. In order to obtain the converged absorption spectrum, we use a uniform $\Gamma$ centred $48 \times 48 \times 1$ grod with a total of 600 bands. We include the top eight valence and bottom eight conduction bands in constructing the kernel matrix. A plane wave cutoff of 25 Ry and 2 Ry is used for the exchange and the screened Coulomb part of the kernel (please look at at Fig S2 for convergence plots). We use the `Elemental` library [S18] to diagonalize the entire BSE Hamiltonian.

Finally, we employ Eqs. (2), (S1) and (S2) to compute the Raman scattering matrix elements from the dipoles, electron-phonon matrix elements and BSE envelope wave-functions. The summation in the Raman scattering matrix element is performed over the first 48000 excitonic states. We define the Raman intensity as the differential cross section, averaged over the polarization of the incoming light and summed over the polarization of the outgoing light.

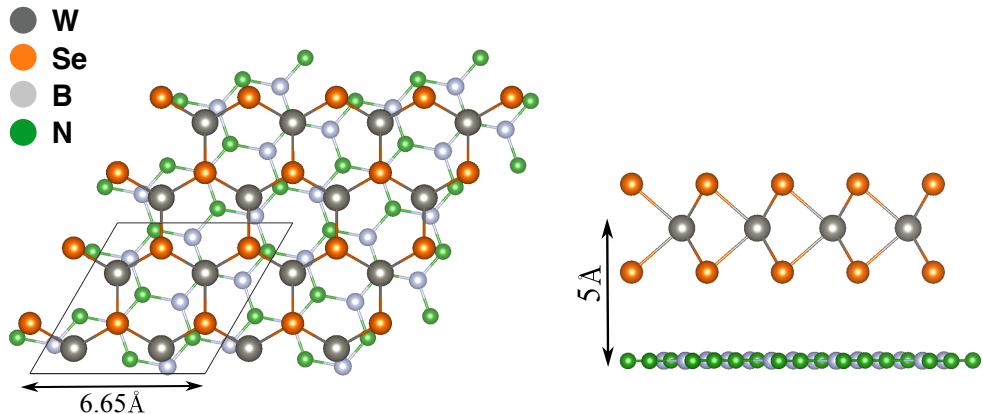

FIG. S1. Top and side view of crystal structure of monolayer WSe$_2$ on top of single layer $h$BN used in the *ab-initio* calculations. Figures were created with the `VESTA` software [S19]

## Convergence parameters

In this subsection, we provide the convergence studies for the BSE and GW calculations with respective to the `YAMBO` input variables as shown in table S1 and Fig. S2.

| YAMBO variable | Meaning | Value |
|---|---|---|
| FFTGvecs | Energy cutoff for expanding the wavefunctions (determines FFT grid for the wavefunctions) | 30 Ry |
| BSENGexx | Energy cutoff for G-vectors used in the construction of exchange part of the BSE kernel, | 30 Ry |
| BSENGBlk | Energy cutoff for G-vectors used in the construction of Screened coulomb potential | 2Ry |
| NGsBlkXs | Energy cutoff for G-vectors used in the construction of dielectric screening matrix | 2Ry |

TABLE S1. Convergence parameters used in the `YAMBO` code. The parameters used to obtain the main results are given in the value column.

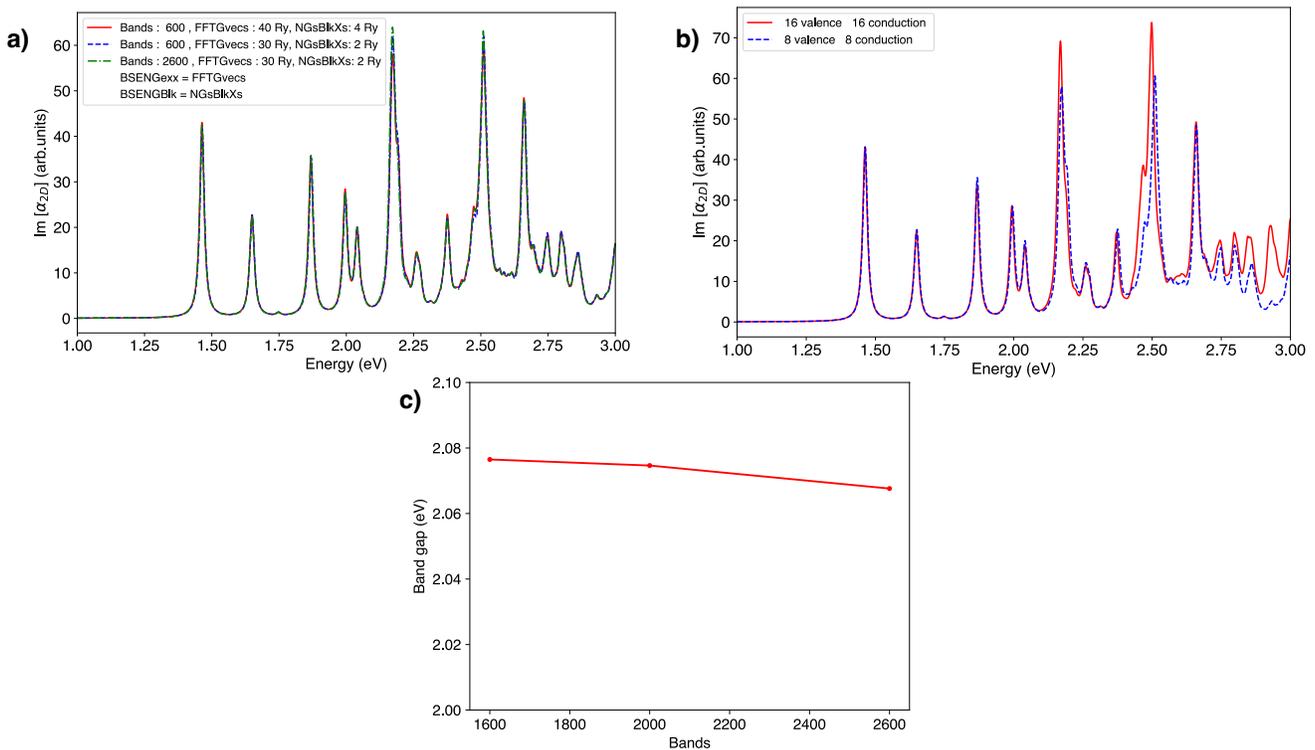

FIG. S2. Convergence of the imaginary part of the 2D polarizability (panels (a) and (b)) and of the $G_0W_0$ bandgap (panel (c)) as a function of different `YAMBO` input parameters. We refer to table. S1 for the meaning of each variable. *"Bands"* in plots (a) and (c) refers to the number of total bands used in the summation, when computing the dielectric tensor within the random phase approximation and in (b) *"valence"* and *"conduction"* refer to the number of valence and conduction bands used in constructing the BSE kernel. For simplicity, these convergence tests are performed on a $12 \times 12 \times 1$ grid while the final calculations have been performed on a $48 \times 48 \times 1$ grid.

# GW BAND STRUCTURE AND EXCITON WAVE FUNCTIONS

In Fig. S3, we show the GW band structure for monolayer WSe$_2$ placed over a single layer of $h$BN. We interpolated the GW corrections on the high symmetry path using the `Wannier90` code

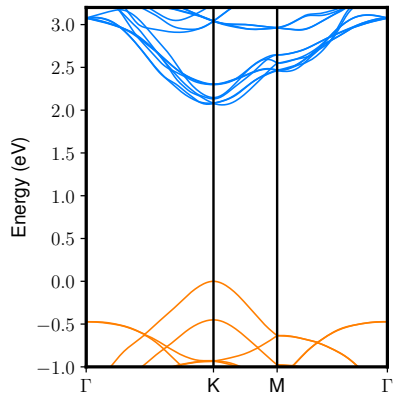

FIG. S3. GW band structure of WSe$_2$/$h$BN heterostructure used in the main paper

In Fig. S4, we plot the excitonic wave functions in reciprocal space ("envelope wave function") defined as

$$\Psi^S(\mathbf{k}) := \sum_{cv} \left| A^S_{\mathbf{k}cv} \right|^2. \tag{S19}$$

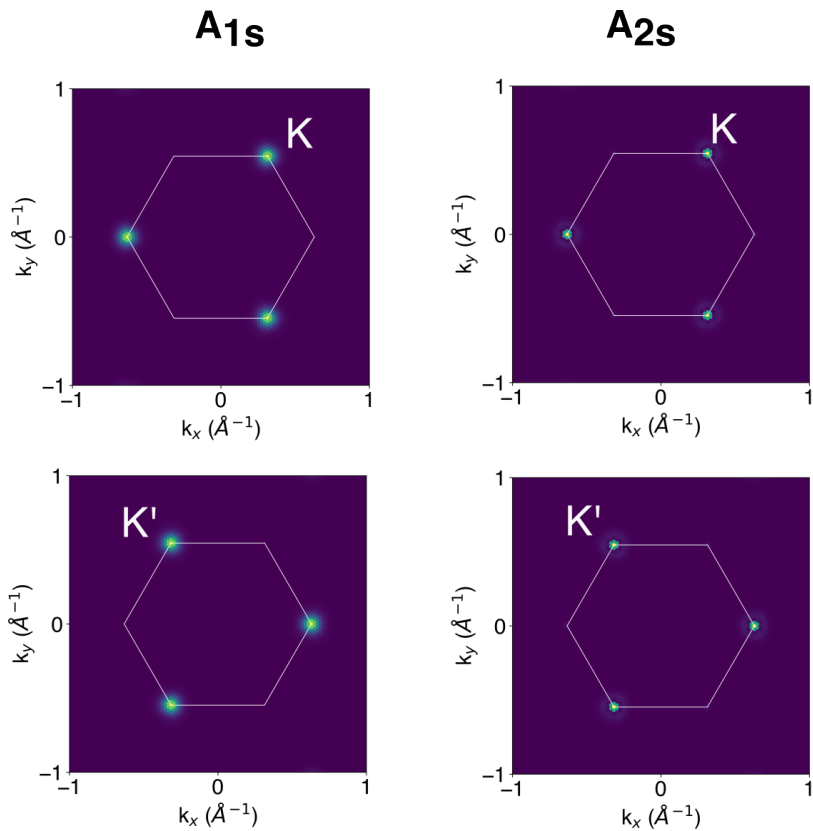

FIG. S4. Envelope wavefunctions for the 1s and the 2s excitons located at two inequivalent K valleys

# DISTANCE DEPENDENCE OF INTERLAYER EXCITON-PHONON COUPLING

Our analysis of inter-layer exciton-phonon interaction suggests that it is highly sensitive to the interlayer distance, as increasing the separation between layers decreases the extent of the valence band hybridization exponentially. To qualitatively validate this, we computed the resonant Raman intensities for the same heterostructure with two different interlayer distances of 5 Å and 6 Å, as shown in Fig. S5. The Raman intensities decrease by two orders of magnitude when the interlayer spacing is increased by 20%.

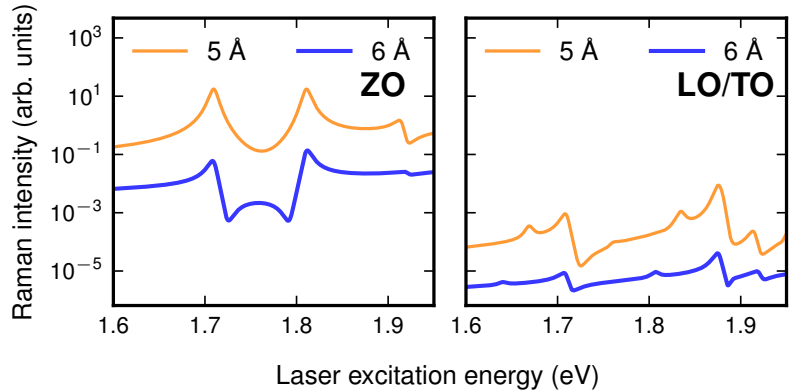

FIG. S5. Resonant Raman intensities of ZO (left) and LO/TO (right) phonons of $h$BN at an interlayer spacing (between tungsten and $h$BN layer) of 5 Å (orange) and 6 Å (blue)

# EXPERIMENTAL DATA

The experimental Raman intensities for the $h$BN ZO mode were obtained by analyzing the raw data recorded for a $h$BN-encapsulated monolayer WSe$_2$ heterostructure, which was provided in the supplementary information of Ref. [S20]. While the experimental Raman spectrum features multiple Raman peaks between 750-820 cm$^{-1}$, we take the intensities of the peak near 815 cm$^{-1}$ as the ZO mode intensities.


\fi

\end{document}